\documentclass[conference]{IEEEtran}
\IEEEoverridecommandlockouts
\usepackage{cite}
\usepackage{amsmath,amssymb,amsfonts}
\usepackage{algorithmic}
\usepackage{graphicx}
\usepackage{textcomp}
\usepackage{xcolor}
\usepackage{float} 
\usepackage{booktabs} 
\usepackage{multirow} 
\usepackage{caption} 
\usepackage{graphicx} 
\usepackage{balance}

\def\BibTeX{{\rm B\kern-.05em{\sc i\kern-.025em b}\kern-.08em
    T\kern-.1667em\lower.7ex\hbox{E}\kern-.125emX}}
\begin{document}
\title{Multi-modal Speech Enhancement with Limited Electromyography Channels\\
}

\author{\IEEEauthorblockN{Fuyuan Feng$^1$, Longting Xu$^1$ and Rohan Kumar Das$^2$}
\IEEEauthorblockA{\textit{$^1$College of Information Science and Technology, Donghua University, Shanghai, China}}
\IEEEauthorblockA{\textit{$^{2}$Fortemedia Singapore, Singapore}\\
2232069@mail.edu.dhu.cn, xlt@dhu.edu.cn, rohankd@fortemedia.com}
}

\maketitle
\begin{abstract}
Speech enhancement (SE) aims to improve the clarity, intelligibility, and quality of speech signals for various speech enabled applications. However, air-conducted (AC) speech is highly susceptible to ambient noise, particularly in low signal-to-noise ratio (SNR) and non-stationary noise environments. Incorporating multi-modal information has shown promise in enhancing speech in such challenging scenarios. Electromyography (EMG) signals, which capture muscle activity during speech production, offer noise-resistant properties beneficial for SE in adverse conditions. Most previous EMG-based SE methods required 35 EMG channels, limiting their practicality. To address this, we propose a novel method that considers only 8-channel EMG signals with acoustic signals using a modified SEMamba network with added cross-modality modules. Our experiments demonstrate substantial improvements in speech quality and intelligibility over traditional approaches, especially in extremely low SNR settings. Notably, compared to the SE (AC) approach, our method achieves a significant PESQ gain of 0.235 under matched low SNR conditions and 0.527 under mismatched conditions, highlighting its robustness.
\end{abstract}


\vspace{-4mm}
\section{Introduction}
Speech enhancement (SE) aims to improve the intelligibility and quality of noisy speech signals, which is crucial for applications like speech recognition and hearing aids. Traditional methods, such as spectral subtraction, Wiener filtering and nonnegative matrix factorization (NMF) \cite{xu2021speech} often struggle in complex noise environments. The rise of deep learning has greatly advanced SE, with early models like feedforward neural networks (FNNs), convolutional neural networks (CNNs) \cite{fu2016snr,li2020speech}, and long short-term memory (LSTM) networks \cite{weninger2015speech,liang2020real,chen2015integration} focusing on the time-frequency domain to estimate clean speech. More recently, end-to-end architectures \cite{fu2018end} have been developed to directly estimate clean speech in the time domain, achieving better results. Advanced approaches, such as generative adversarial networks (GANs) \cite{abdulatif2024cmgan,donahue2018exploring,phan2020improving,ji2020adversarial} and attention-based models, have further enhanced SE. 

A recent promising development in SE is the Mamba architecture, which leverages state-space models with a selection mechanism. Rong Chao et al. \cite{chao2024investigation} conducted a comparative study of transformer-based and Mamba-based SE models and introduced a novel system namely SEMamba. Their approach utilizes a bidirectional Mamba architecture where the input is processed in parallel through the Mamba network and subsequently connected to the output. This bidirectional Mamba is applied concurrently in both the time and frequency domains, a configuration referred to as TF-Mamba. Their studies demonstrate that TF-Mamba shows significant potential for improving SE performance.

\begin{figure*}[h] 
    \centering
    \includegraphics[width=\textwidth]{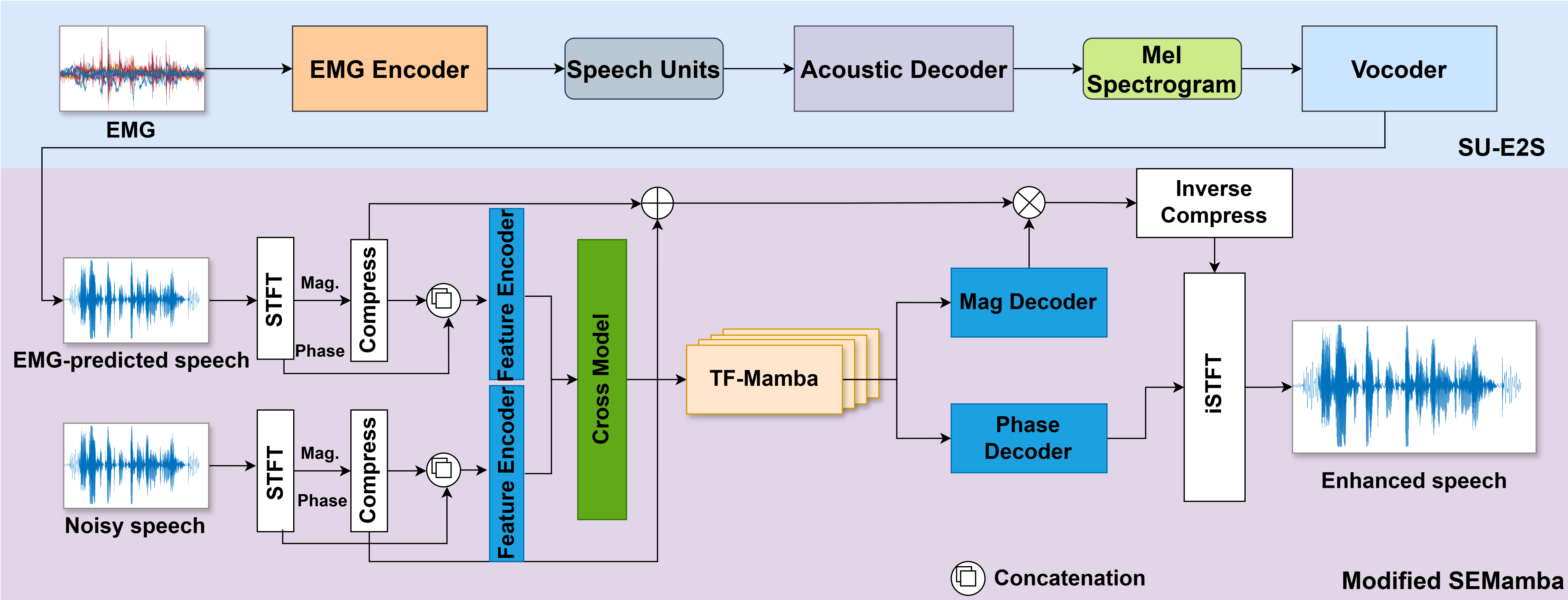} 
    \caption{Architecture of multi-modal SE based on SU-E2S and modified SEMamba.} 
    \label{fig:emg-se} 
\end{figure*}

All these mentioned SE methods were developed using air-conducted (AC) speech. However, due to the nature of air conduction, AC speech is highly susceptible to ambient noise, which significantly degrades its performance in low SNR and non-stationary noise environments. For example, security guards or on-site agents who perform tasks at famous sports events will have a lot of noise, especially language noise~\cite{taherian2020robust}. To overcome these limitations, alternative modalities like video~\cite{hou2018audio} have been explored to enhance target speech.

However, audio-visual enhancement relies on camera-equipped devices, limiting its use in outdoor, underwater, or military settings. Another promising solution is the use of Electromyography (EMG) signals, which can be non-invasively recorded through skin-attached electrodes. EMG signals from the throat and face during speech provide valuable speech-related information, and numerous studies have shown the viability of EMG in speech applications.

Diener et al. \cite{diener2016initial} compute stacked time-domain features from windowed EMG signals and map these features to parallel acoustic features, subsequently, a vocoder synthesizes the acoustic waveform from the acoustic feature predictions. Gaddy and Klein \cite{gaddy2020digital,gonzalez2016silent} process EMG signals with convolutional layers and a Transformer encoder. They align EMG signals of silent articulation with a reference audio by performing dynamic time warping (DTW) \cite{muda2010voice}.
Matthias Janke et al. \cite{janke2017emg} proposed directly converting EMG signals into audible speech waveforms and compared methods for extracting time-domain features from EMG signals. 

In~\cite{scheck2023multi}, the authors proposed Speech-Unit-based EMG-to-Speech (SU-E2S) modal, which predicts soft speech units from EMG signals and uses a pre-trained acoustic VC decoder to reconstruct acoustic features. Inspired by this article, SU-E2S modal was also adopted in this work due to the fact that the EMG encoder of the SU-E2S system is not trained to predict the original acoustic features, but rather representations of the spoken content. Therefore, we hope to learn richer semantic information independent of speaker features~\cite{Scheck24EMBC}. Recently, the authors of~\cite{wang2022emgse} applied EMG signals to SE for the first time, using noisy spectrogram and 15 time-domain features (TD15 vectors) extracted from 35 channels of EMG as inputs for SE, and achieved improved results in some noisy scenarios. 

However, previous EMG-based SE methods typically required 35 EMG channels, which significantly limited their usage for practical applications. In contrast, our approach is designed to use only 8 EMG channels, making it more feasible for real-world scenarios. To achieve this, we have adapted the latest SEMamba framework for multi-modal SE by integrating additional input modalities to enhance the performance. Our approach operates in two stages: In the first stage, we predict speech signals from 8-channel EMG signals that are unaffected by environmental noise. In the second stage, these predicted speech signals, together with noisy speech signals, are used as inputs to our model to further enhance the speech. This two-stage strategy contributes in both speech quality and intelligibility compared to traditional noise-only methods, especially in cases with extremely low SNR.
 
\section{Multi-Modal SE: Processing AC Speech and EMG Signals with SEMamba}

As illustrated in Fig.~\ref{fig:emg-se}, the proposed multi-modal SE based on the modified SEMamba has two stages. In the first stage, EMG signals recorded during voiced pronunciation are converted into speech signals. Then, in the second stage, the predicted speech from EMG and the noisy speech are taken as input into our proposed multi-modal SE. The details of the two stages are described in the following subsections. 

\subsection{Speech-Unit-based EMG-to-Speech (SU-E2S)}

For the EMG encoder, we adopt the network architecture from Gaddy and Klein \cite{gaddy2020digital}, incorporating modifications specifically for soft SU prediction. Initially, a convolutional layer downsamples the EMG signal \((\mathbf{X}_1, \dots, \mathbf{X}_{T})\) to a frame rate of 50Hz, aligning it with the frequency of the soft speech units. The resulting feature sequence is then processed by a transformer encoder, which outputs two predictions: the soft speech units \((\hat{c}_1, \ldots, \hat{c}_{\hat{C}})\) and the phonemes \((\hat{p}_1, \ldots, \hat{p}_{\hat{C}})\). The encoder is trained by minimizing the distance between the predicted and target soft speech units \(({c}_1, \ldots, {c}_{{C}})\), as well as by optimizing the cross-entropy loss for phoneme classification, a method shown to enhance performance in related studies.
\begin{equation}
\mathcal{L}_{SU} = \frac{1}{C} \sum_{t=1}^{C} \left\| \mathbf{c}_t - E_c(\mathbf{x})_{t}\right\|_2
\end{equation}
where \(E_c(\mathbf{x})_t\) denotes the speech units predictions of the EMG encoder at frame $t$.
\begin{equation}
\mathcal{L}_{P} = -\frac{1}{C} \sum_{t=1}^{C} \sum_{i=1}^{|\mathcal{P}|} \mathbf{b}_{t,i} \cdot \log E_p(\mathbf{x})_{t,i}
\end{equation}
where \(\mathbf{b}_{t,i}\) is the binary indicator that phoneme \(i\) is the target class at frame \(t\).  \(E_p(\mathbf{x})_t\) is a sequence of probability distributions for phoneme predictions, with a length of \(C\).
The total loss is a weighted sum of \(\mathcal{L}_{SU}\) and \(\mathcal{L}_{P}\):
\begin{equation}
\mathcal{L}_{\text{total}} = \lambda_{SU} \mathcal{L}_{SU} + \lambda_{P} \mathcal{L}_{P}
\end{equation}
Where \(\lambda_{SU}\) and \(\lambda_{P}\) are scalar weights for the respective components.

After obtaining the predicted soft speech units \cite{scheck2023multi}, we use a pre-trained acoustic decoder with a convolutional module \cite{van2022comparison}, a Pre-Net, and an autoregressive LSTM to convert predicted speech units into Mel-spectrograms. The network's encoder-decoder structure transforms discrete or soft speech units into spectrograms for efficient voice conversion. 
Subsequently, just like the direct EMG-to-Speech model \cite{diener2016initial,scheck2023stream}, a vocoder synthesizes the acoustic signal from the predicted features. Here, we use a pre-trained HiFi-GAN model \cite{kong2020hifi} to perform this synthesis, ensuring high-quality audio generation from the predicted Mel spectrograms.

This is the process of the first stage, highlighted in the blue box at the top of Fig.~\ref{fig:emg-se}, involves converting the EMG signal into an EMG-predicted speech signal that is robust to environmental noise by predicting soft speech units.

\subsection{Multi-modal SE based on modified SEMamba}

Our implementation of multi-modal SE is based on SEMamba \cite{chao2024investigation}. In environments with excessive noise, enhancing the noisy speech signals alone may not produce satisfactory results \cite{hou2018audio,tagliasacchi2020seanet}. Therefore, we use both the EMG-predicted speech, obtained as described in the previous subsection, and the noisy speech as inputs. These two signals are synchronized and correspond to the same clean speech.

As depicted in the purple box in the lower half of Fig.~\ref{fig:emg-se}, the process begins with applying short-time Fourier transform (STFT) to both input signals to obtain their spectral representations. The magnitude component is then compressed and stacked with the phase component. These stacked components are fed into a feature encoder, which performs initial feature extraction on each of the speech signals separately. The feature encoder uses a DenseEncoder architecture with a DenseNet core featuring dilated convolutions, complemented by standard convolutional layers on either end for multi-scale feature extraction. Following this initial extraction, a cross module comprising fully connected layers is employed to fuse the features from both signals, effectively integrating rich speech information from each source. The fused output is then processed by the TF-Mamba module, which extracts deeper time-frequency domain characteristics.

TF-Mamba consists of two bidirectional Mamba blocks operating in both the time and frequency domains. The input is processed in parallel through the Mamba network, and the outputs are concatenated. This combined output is then fed into a \textit{Conv1D} layer, formulated as:
\begin{equation}
\mathbf{y} = \textit{Conv1D}(M_{\mathit{uni}}(\mathbf{x}) \oplus M_{\mathit{uni}}(\mathit{flip}(\mathbf{x})))
\end{equation}
where  \(\mathbf{x}\), \(\mathbf{y}\), \(M_{\mathit{uni}}()\), \(\mathit{flip}()\), \textit{Conv1D}(), and \(\oplus\) represent the input, output, uni-directional Mamba operation, flipping operation, 1-D convolution, and concatenation, respectively.

The output from TF-Mamba is then fed into two separate decoders: one reconstructs the magnitude mask, while the other reconstructs the real and imaginary parts of the waveform. Both decoders use DenseBlock structures with dilated convolutions and convolutional layers to facilitate feature extraction as well as reconstruction. Finally, the enhanced speech is obtained by applying the inverse short-time Fourier transform (iSTFT).

\section{Experiments}

\subsection{Dataset}\label{AA}
Our study focuses on speech and EMG signals in scenarios involving audible speech production. Therefore, we evaluate our model on the corpus from Gaddy and Klein \cite{gaddy2020digital}, which documents subjects reading English sentences under both audible and silent articulation conditions. The EMG signals in this corpus are recorded in 8-channel with a sampling rate of 1000Hz, while the audio are recorded in 16 kHz. We use the same EMG filtering steps as in the authors’ implementation in~\cite{scheck2023multi}. We use the predefined validation and testing splits, but use utterances with EMG signals of audible articulation. Then the train, validation, and test splits contain 6755, 199, and 98 utterances, respectively.
For the training and validation sets, we applied MUSAN~\cite{snyder2015musan} to generate noisy audio data. Each utterance is corrupted with five randomly selected types of noise at five SNRs (-10, -5, 0, 5, and 10 dB). For the test set, we used 18 unseen noise types (car noise, engine noise, pink noise, white noise, two types of street noises, six background Chinese speakers, and six English speakers) to clean utterances at four SNRs (-11, -6, -1, and 4 dB) to create mismatch conditions~\cite{wang2022emgse}. It is noted that background speakers in different languages are used to create babble noise in language-specific conditions. 

\begin{table*}[t!]
\centering
\caption{Performance of SE (AC) and proposed method with different number of TF-Mamba blocks under different SNR. The matched condition uses 5 noises from MUSAN at 4 SNRs (-10, -5, 0, 5 dB) similar to training, while the mismatched condition uses 18 unseen noises at SNRs (-11, -6, -1, 4 dB).}
\resizebox{\textwidth}{!}{ 
\begin{tabular}{c c cc cc cc cc cc cc cc}
\toprule
\multirow{2}{*}{} & \multirow{2}{*}{\textbf{SNR}} & \multicolumn{2}{c}{\textbf{Noisy speech}} & & \multicolumn{2}{c}{\textbf{SE (AC)}} & & \multicolumn{2}{c}{\textbf{Proposed (TF-Mamba$\times$1)}}& & \multicolumn{2}{c}{\textbf{Proposed (TF-Mamba$\times$4)}} & & \multicolumn{2}{c}{\textbf{Proposed (TF-Mamba$\times$8)}}\\
\cmidrule{3-4} \cmidrule{6-7} \cmidrule{9-10} \cmidrule{12-13} \cmidrule{15-16}
 & & \textbf{PESQ} & \textbf{STOI} & & \textbf{PESQ} & \textbf{STOI} & & \textbf{PESQ} & \textbf{STOI} & & \textbf{PESQ} & \textbf{STOI} & & \textbf{PESQ} & \textbf{STOI}\\
\midrule
\multirow{4}{*}{\textbf{Match}} 
& -10db & 1.304 & 0.682 & & 2.641 & 0.898 & & 2.672 & 0.905 & & \textbf{2.876} & \textbf{0.925} & & 2.778 & 0.917 \\
& -5db  & 1.26  & 0.75  & & 2.945 & 0.929 & & 2.9   & 0.928 & & \textbf{3.094} & \textbf{0.943} & & 3.024 & 0.937 \\
& 0db   & 1.365 & 0.831 & & 3.327 & 0.954 & & 3.24  & 0.95  & & 3.404 & 0.961 & & 3.363 & 0.956 \\
& 5db   & 1.54  & 0.89  & & 3.624 & 0.971 & & 3.519 & 0.967 & & 3.655 & 0.974 & & 3.624 & 0.97 \\
\cmidrule(lr){2-16} 
& average & 1.367 & 0.788 & & 3.134 & 0.938 & & 3.083 & 0.938 & & \textbf{3.257} & \textbf{0.951} & & 3.197 & 0.945\\

\midrule
\multirow{4}{*}{\textbf{Mismatch}} 
& -11db  & 1.174 & 0.533 & & 1.554 & 0.756 & & 1.917 & 0.835 & & \textbf{2.081} & \textbf{0.862} & & 1.972 & 0.851 \\
& -6db  & 1.139 & 0.637 & & 1.885 & 0.846 & & 2.236 & 0.878 & & \textbf{2.444} & \textbf{0.903} & & 2.353 & 0.893\\
& -1db & 1.176 & 0.738 & & 2.263 & 0.901 & & 2.544 & 0.91 & & 2.774 & 0.931 & & 2.712 & 0.923 \\
& 4db & 1.262 & 0.831 & & 2.77 & 0.935 & & 2.91 & 0.938 & & 3.126 & 0.954 & & 3.083 & 0.947 \\
\cmidrule(lr){2-16} 
& average & 1.188 & 0.685 & & 2.118 & 0.86 & & 2.402 & 0.89 & & \textbf{2.606} & \textbf{0.913} & & 2.53 & 0.904\\
\bottomrule
\end{tabular}
}
\label{tab1}
\end{table*}

\subsection{Reference baselines and proposed method}

In our experiments, we compared the performance of our proposed method against several baselines to evaluate its effectiveness in different noise environments, which are:
\begin{itemize}
\item Noisy Speech: This baseline represents the raw, unprocessed noisy speech. It serves as the starting point for quality and intelligibility without any enhancement.
\item SE (AC): It has the same network structure to that of TF-Mamba$\times$4, with the difference being that the input only considers noisy speech but no EMG signals, thus developed for uni-modal SE.
\item Proposed: It is the multi-modal SE based on modified SEMamba. We tested with different numbers of TF-Mamba blocks, including 1, 4, and 8 blocks, referred to as TF-Mamba$\times$1, TF-Mamba$\times$4, and TF-Mamba$\times$8, respectively. Each configuration utilizes both EMG and noisy speech to improve the quality and comprehensibility of speech.
\end{itemize}

\subsection{Implementation details and evaluation metrics}
In the first stage, we train the EMG encoder by minimizing the loss function $\mathcal{L}_{\text{total}}$ with weights \(\lambda_{SU}\) = 0.5 and \(\lambda_{P}\) = 0.5, and the learning rate is set to 0.0003. For the pre-trained acoustic decoder, we use a learning rate of 0.0001 and learn for 80k steps. The loss function in the second stage is a linear combination of the PESQ-based GAN discriminator loss, and losses based on time, magnitude, complex, and phase components~\cite{lu2023mp}.

In this study, we use the perceptual
evaluation of speech quality (PESQ)~\cite{rix2001perceptual} as an objective quality measure to assess the quality of the speech signals. In addition, short-time objective intelligibility (STOI)~\cite{taal2011algorithm} is used for intelligibility assessment of the speech signals. 

\section{Results and Discussion}

Table~\ref{tab1} shows the performance comparison of our proposed method to the reference baselines under various conditions. We define the matched condition as the test set comprising 5 noises randomly selected from MUSAN~\cite{snyder2015musan} at 4 SNRs (-10, -5, 0, and 5 dB), consistent with the types of noise encountered during training. In contrast, the mismatched condition uses a test set of 5 noises randomly selected from 18 unseen noises at SNRs of -11, -6, -1, and 4 dB. We discuss the results of various studies conducted in the following subsections. 

\subsubsection{Matched low SNR conditions}
Under matched conditions (-10 dB and -5 dB), TF-Mamba$\times$4 demonstrates substantial improvements over the baseline SE (AC) method. At -10dB, TF-Mamba$\times$4 achieves a PESQ score of 2.876, compared to SE (AC)’s 2.641, marking an improvement of 0.235. Similarly, at -5dB, TF-Mamba$\times$4 obtains a PESQ of 3.094, exceeding SE (AC) by 0.149. These gains are particularly significant given the difficult nature of these low SNR environments. The STOI scores also show a similar pattern, with TF-Mamba$\times$4 achieving 0.925 at -10dB and 0.943 at -5dB, indicating clear enhancements in speech intelligibility.

\subsubsection{Mismatched low SNR conditions}
In mismatched conditions (-11 dB and -6 dB), the benefits of TF-Mamba$\times$4 are even more pronounced. At -11dB, TF-Mamba$\times$4 achieves a PESQ of 2.081, significantly higher than SE (AC)’s 1.554, representing an increase of 0.527. This enhancement persists at -6 dB, where TF-Mamba$\times$4 reaches a PESQ score of 2.444, outperforming the baseline’s 1.885 by 0.559. The STOI improvements are similarly notable, with TF-Mamba$\times$4 achieving 0.862 at -11dB and 0.903 at -6dB, reflecting substantial gains in intelligibility under these extreme conditions.

\subsubsection{Ablation study with different number of TF-Mamba blocks}
Table~\ref{tab1} also demonstrate that increasing the number of TF-Mamba blocks from 1 to 4 leads to significant improvements in performance, while further increasing to 8 blocks results in only marginal gains, indicating diminishing returns. The minimal difference between TF-Mamba$\times$4 and TF-Mamba$\times$8 suggests that four blocks strike an optimal balance between performance and computational efficiency.

We are now interested to observe the impact of proposed method under different types of noise. Table~\ref{tab2} highlights the effectiveness of our TF-Mamba$\times$4 method compared to the baseline noisy speech and SE (AC) approaches across various noise types. Our method consistently outperforms the SE (AC) in both PESQ and STOI scores. Notably, in complex noise environments like pink and white noise, it shows greater improvements in speech quality and intelligibility. Again, considering the babble noise in English and Chinese speaking speaker environments, the relative performance improvement was more significant in English babble noise environments.   



\begin{table}[t!]
\centering
\caption{Performance of SE (AC) and proposed method (TF-Mamba$\times$4) for different noise types.}
\resizebox{\columnwidth}{!}{ 
\begin{tabular}{c cc cc cc cc}
\toprule
\multirow{2}{*}{\textbf{}} & \multicolumn{2}{c}{\textbf{Noisy speech}} & & \multicolumn{2}{c}{\textbf{SE (AC)}} & & \multicolumn{2}{c}{\textbf{Proposed} } \\
\cmidrule{2-3} \cmidrule{5-6} \cmidrule{8-9}
 & \textbf{PESQ} & \textbf{STOI} & & \textbf{PESQ} & \textbf{STOI} & & \textbf{PESQ} & \textbf{STOI} \\
\midrule
Car & 1.117 & 0.762 & &  2.351 & 0.903 & &  2.588 & 0.926 \\
Engine & 1.086 & 0.645 & &  1.821 & 0.793 & &  2.277 & 0.873 \\
Pink & 1.056 & 0.69 & &  1.73 & 0.805 & &  \textbf{2.233} & \textbf{0.881} \\
White & 1.051 & 0.729 & &  1.785 & 0.832 & &  2.281 & 0.894 \\
Street & 1.145 & 0.673 & &  1.979 & 0.83 & &  2.386 & 0.891 \\
English  & 1.213 & 0.665 & &  2.121 & 0.869 & &  \textbf{2.707} & \textbf{0.92} \\
Chinese & 1.247 & 0.694 & &  2.312 & 0.878 & &  \textbf{2.758} & \textbf{0.925} \\
\bottomrule
\end{tabular}
}
\label{tab2}
\end{table}

These results discussed above for various noise conditions and types demonstrate that our proposed TF-Mamba$\times$4 model effectively integrates multi-modal information from EMG and speech, leading to substantial improvements in speech quality and intelligibility in challenging and language-specific babble noise conditions. It is worth to be noted that this was achieved by considering only 8 channels of EMG signals.

\section{Conclusion}
In this study, we proposed a modified SEMamba framework for multi-modal SE using 8-channel EMG signals combined with noisy speech. This approach is evaluated in challenging environments with low SNR and language-specific babble noise conditions. The integration of EMG with noisy speech outperformed the uni-modal method that only considers noisy speech, showing a significant improvement in PESQ and STOI. Our analysis suggest that using four TF-Mamba blocks provides an optimal balance between performance and computational efficiency, with limited gains beyond this point. Notably, our method employs only 8 EMG channels, the lowest count in multi-modal speech enhancement studies considering EMG signals, underscoring its efficiency. However, there are challenges remain in effectively fusing features from different modalities, which may limit enhancement performance. Future research will focus on developing advanced feature fusion techniques to fully leverage the complementary information from EMG and acoustic signals, aiming to enhance SE performance in diverse real-world environments.  

\clearpage

\balance
\bibliographystyle{IEEEtran}
\bibliography{strings}
\vspace{12pt}
\end{document}